# Constructing experimental indicators for Open Access documents

Philipp Mayr

Doctoral candidate at the Institute of Library and Information Science,
Humboldt-University Berlin, Germany.

E-mail: philippmayr@web.de
http://www.ib.hu-berlin.de/~mayr

Abstract

The ongoing paradigm change in the scholarly publication system ('science is turning to e-science') makes it necessary to construct alternative evaluation criteria/metrics which appropriately take into account the unique characteristics of electronic publications and other research output in digital formats. Today, major parts of scholarly Open Access (OA) publications and the self-archiving area are not well covered in the traditional citation and indexing databases. The growing share and importance of freely accessible research output demands new approaches/metrics for measuring and for evaluating of these new types of scientific publications. In this paper we propose a simple quantitative method which establishes indicators by measuring the access/download pattern of OA documents and other web entities of a single web server. The experimental indicators (search engine, backlink and direct access indicator) are constructed based on standard local web usage data. This new type of web-based indicator is developed to model the specific demand for better study/evaluation of the accessibility, visibility and interlinking of open accessible documents. We conclude that e-science will need new stable e-indicators.

## 1. Introduction

Web data driven studies are a relatively new, interdisciplinary and demanding research field in the information science sector. The impetus for this mostly empirical oriented research comes from the rampant increase in users, documents and scholarly research output on the Web. In these studies there are two types of essential data: Web link data, well-explored by Mike Thelwall (Thelwall, 2004) and other researchers; and the relatively under-valued web usage data, which we have been studying.

Web log files (web usage data) record user transactions on web servers and other openly accessible document repositories and provide an excellent source for information behavior analysis due to their properties and amount of data (Nicholas et al., 2005a; 2005b). In addition to this well-known analysis path, usage data offer the possibility to draw conclusions concerning accessibility and visibility including interlinking of various types of documents. Recent informetric studies demonstrate the potential to build indicators on this kind of data (Brody & Harnad, 2005; Bollen et al., 2005).

Our approach is also based on web usage data. This approach could be relevant for the field of research evaluation because in the electronic age of publishing science, policy makers and reviewers, as well as others who use these results, need proper instruments. There is a need to construct new metrics which are comparable or compatible to citation impact. Perhaps measuring usage impact or new metrics of user preference can be a supplement for citation impact.

"Is it possible to identify the importance of an article earlier in the read-cite cycle, at the point when authors are accessing the literature? Now that researchers access and read articles through the Web, every download of an article can be logged. The number of downloads of an article is an indicator of its usage impact, which can be measured much earlier in the reading-citing cycle." (Brody & Harnad, 2005)

At a first step, the paper introduces to the field of Open Access (OA) and webometrics, including web-based indicators. Secondly we propose three experimental web indicators and shortly describe their construction. The paper ends with a discussion and a conclusion of the relevance and specific value of the proposed indicators for research evaluation.

## 2. Open Access and Web indicators

According to advocates of OA and recent political declarations (e.g. Berlin declaration, 2003), the share and impact of OA journals as well as the self-archiving area (Lawrence, 2001) will extensively increase in coming years. In fact, this OA sector is still underrepresented and not well covered by the major indexing services (e.g. McVeigh, 2004). The consequence is a lack of established instruments to measure or evaluate the OA published and archived research output. All stakeholders involved in the scientific process (scientists, peers, journal editors, librarians, etc.) have indicated that robust web data based measuring methods and indicators[1], e.g. for determining the impact or quality of a document, are still lacking. There are presently no sustainable and simple standard indicators in sight. Only experimental approaches have been proposed (web link indicators included).
Open Access repositories like CiteSeer, ArXiv and others are especially suited for developing web based indicators because they are defined by their scope and offer a certain degree of import control and content quality. The growing share and importance of freely accessible research documents in the scholarly communication demands robust web-based indicators now. Otherwise evaluation of these new publication types is not feasible.

The WISER project defines a web indicator as "a policy relevant measure that quantifies aspects of the creation, dissemination and application of science and technology in so far as they are represented on the internet or the World Wide Web." (see WISER project web site http://www.wiserweb.org, 2005).
Current web indicators are not without systematic problems. Webometric research can show, for example, that analyses based on commercial search engine results such as Google APIs have some limitations (Mayr & Tosques, 2005)[2]. Additionally, data from different search engines is not easily compared and interpreted (Bar-Ilan, 1998/9; 2005). Link analysis, intensively investigated by Mike Thelwall (Thelwall 2004), including co-link analysis (Katz, 2004), is one leading direction for future web indicators.
There is perhaps another direction for constructing reliable and comprehensive web-based indicators. The importance of web usage data is also becoming more accepted in the informetric community. Recent papers in a special issue on informetrics in Information Processing & Management (vol. 41(6), 2005) and a workshop at Humboldt University in Berlin[3] show various new uses of this data, e.g. alternative metrics of journal impact (Bollen et al., 2005) or deep log analysis of digital journals (Nicholas et al., 2005b). Modeling web usage data for evaluation purposes is quite new (Brody & Harnad, 2005; Kurtz et al., 2005; Bollen et al., 2005) and requires the appropriate environment (e.g. e-print archives like arxiv.org). Preliminary results from this research show a positive correlation between usage data and citation data.

---

[1] Definition by Mike Thelwall: "Statistics about university web sites are, therefore, process indicators (Scharnhorst, 2004). These can be used to provide information to aid policy makers and managers to ensure that web sites are being used effectively as part of the processes of education and research." (Thelwall, 2004, p:228).
[2] Limitations and differences in search engine APIs data (Google APIs and Yahoo! APIs) can be tested with some live demos, available at: http://bsd119.ib.hu-berlin.de/~ft/index_e.html.
[3] See http://www.dini.de/veranstaltung/workshop/oaimpact/programm.php for the workshop programme and presentations.

The following chapter will deal with three experimental indicators constructed out of standard web usage log data.

## 3. Experimental usage indicators

Conventional analysis of usage data focuses on aggregates of use from a macro perspective. Their common performance figures of hit, view and visit are broadly accepted, but they give only a very crude and incomplete picture of the underlying access pattern, such as:

- How much traffic does an entity (site, directory, page) of a web space receive (see Fig. 1)?
- Where are the most important entry pages?
- Where are the users coming from?
- How do users return?

The limitations and advantages of web usage data analysis has recently been reviewed by Jamali *et al* (2005, see also Nicholas et al., 1999).

We can summarize that log files of web servers or OA repositories normally are:
- structured enough to reliable extract pattern and instances of usage,
- can be used for measuring web impact or usage impact (Brody & Harnad, 2005) of a certain entity,
- have the potential for fast indication of usage trends, including user interest and preference.

Below we will propose a simple quantitative method which establishes indicators by measuring the access pattern of selected OA documents and other web entities of a web server (see Fig. 1 for an illustration of the distinguishable entities). The procedure was developed in 2004 and tested by an open accessible academic web site (Mayr, 2004a; 2004b). We will call the measure Web Entry measure.

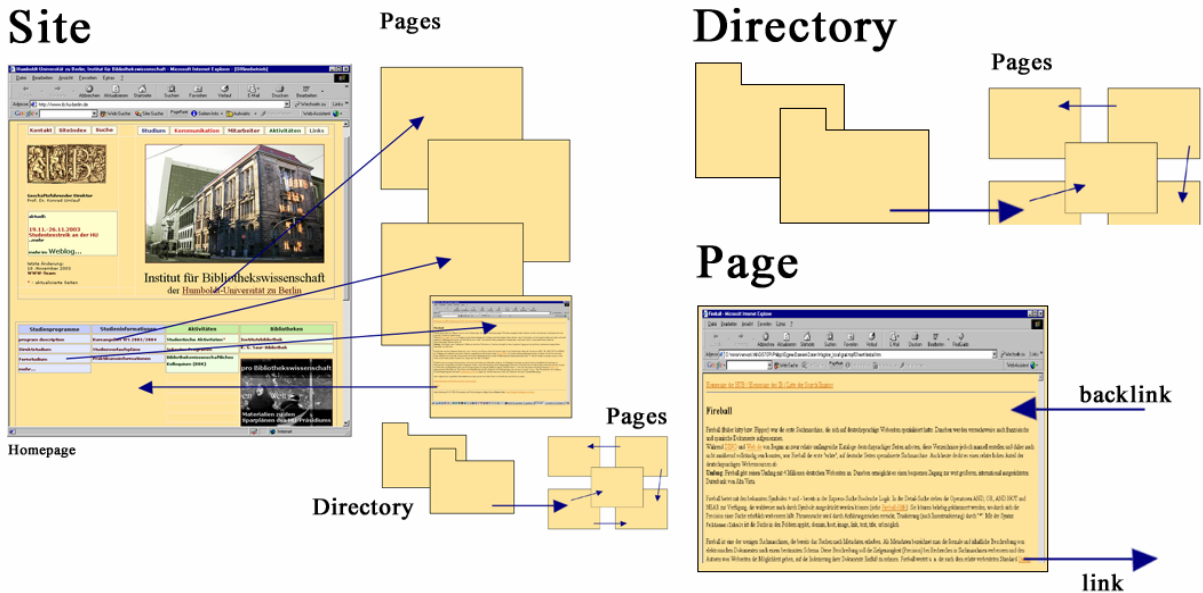

Figure 1: Different web entities (site, directory and page) in an open accessible web space.

The main idea of the approach is to distinguish the access or entry pattern of a document or entity via a simple heuristic. The Web Entry measure uses a heuristic that distinguishes three access/download types "*search engines*," "*backlinks*" and "*direct access*" (see Fig. 2) for measuring the accessibility, interlinking and visibility ratios for certain documents. These access or entry types can be reliably identified in a log file (see also Thelwall, 2001; Mayr, 2004a; see Fig. 2). The usage data in the extended log file format allows a separation of access requests into (see log file examples in Fig. 2):

1. *Backlink* access: the usage of a hyperlink (backlink for the requested item) in another external paper or web document is documented in the usage data with the complete URL of the referencing document.
2. *Search engine* access: a search engine query e.g. the title or the author of a paper, leads directly to the paper in the repository and will be logged as an access via search engine. Search engine access can be tracked by typical search engine urls in the referer log file field.
3. *Direct* access: direct access can be established through user's bookmarks or other forms of directly requesting a document. Direct access can be tracked by the missing entry/url (e.g. "-") in the referer log file field.

```
# Type backlink access
141.20.20.xx - - [20/Jul/2002:22:50:55 +0200] "GET /~wumsta/ubach/fuss.htm HTTP/1.1" 200 54988
"http://www.referrer.com/article11.htm" "Mozilla/4.0 (compatible; MSIE 6.0; Windows NT 5.1)"

# Type search engine access
203.122.23.xxx - - [20/Jul/2002:23:14:37 +0200] "GET /~pbruhn/gruppe04.htm HTTP/1.1" 200 62766
"http://www.google.de/search?q=%2B%22russische+Frauen%22" "Mozilla/4.0 (compatible; MSIE 6.0;
Windows NT 5.0)"

# Type direct access
200.109.102.xxx - - [20/Jul/2002:23:14:38 +0200] "GET /index.html HTTP/1.0" 200 279
"-" "BlitzBOT@tricus.net (Mozilla compatible)"
```

Figure 2: A log file sample showing the proposed access types search engine, backlink and direct access which can be differentiated by a heuristic.

Figure 2 shows anonymous virtual users requesting different pages (bold). The users access the pages via the three proposed entry types (typical search engine, backlink and direct access pattern are marked).

The Web Entry measure proposed by the author can be implemented[4] in a freely accessible environment of Open Access documents. The differentiation of the entry requests allow measuring of, for example, the intensity and broadness of external interlinking or visibility via search services. This has been successfully applied in a study analyzing the content of a large academic website in Germany (Mayr, 2004a; 2004b).

Furthermore, the Web Entry measure has the potential for building fast applicable indicators using these aggregates of usage which may be a basis for various optimization or evaluation processes of a certain OA entity. Detailed usage impact and interest rates can be indicated shortly after electronic publication by the following ratios:

- Web entity visibility or accessibility rates for different navigation and document access pattern,
- Web entity interlinking rates give a global idea about the accessibility of an open accessible item by external linking/backlinking or electronic scholarly citing.

The three entry patterns (exemplified by Fig. 2) can be extracted for entities, for example journals, journal volumes/issues or articles/abstracts etc, identifiable in the usage data. The following simple pure web-based indicators can be automatically computed and used for further calculations (e.g. validations).

1. Search engine Indicator $I_{se}$:

$$I_{se} = \frac{\text{Downloads by search engine access } (D_{se})}{\text{Downloads total } (D_{total})}$$

2. Backlink Indicator $I_{bl}$:

$$I_{bl} = \frac{\text{Downloads by backlink access } (D_{bl})}{\text{Downloads total } (D_{total})}$$

3. Direct access Indicator $I_{da}$:

$$I_{da} = \frac{\text{Downloads by direct access } (D_{da})}{\text{Downloads total } (D_{total})}$$

The entity in Table 1, a single web page (article.htm in the folder irs), is downloaded 10.000 times in a certain period of time. The proposed analysis method subdivides these total downloads into 2.000 downloads via search engines ($I_{se}$= 0.2), 7.000 downloads via backlinks ($I_{bl}$= 0.7) and 1.000 downloads via direct access ($I_{da}$= 0.1). The single indicator values $I_{se}$, $I_{bl}$, $I_{da}$ can vary between 0 and 1. Each indicator can be drilled down to a more detailed level (see 2-step scenario below).

---

[4] The prototype Web Entry Miner (WEM) and sample log data are available at http://www.ib.hu-berlin.de/~mayr/wem/.

Table 1: An example for segmented download ratios and derived indicators $I_{se}$, $I_{bl}$, $I_{da}$ for an entity.

| Entity | URL | $D_{se}$ ($I_{se}$) | $D_{bl}$ ($I_{bl}$) | $D_{da}$ ($I_{da}$) | $D_{total}$ |
|---|---|---|---|---|---|
| Page | /irs/article.htm | 2.000 (0.2) | 7.000 (0.70) | 1.000 (0.10) | 10.000 |

Figure 3 shows a prototypical visualization of usage data via the proposed measure. The values of the selected page indicate a high search engine ratio ($I_{se}$ of 0.75).

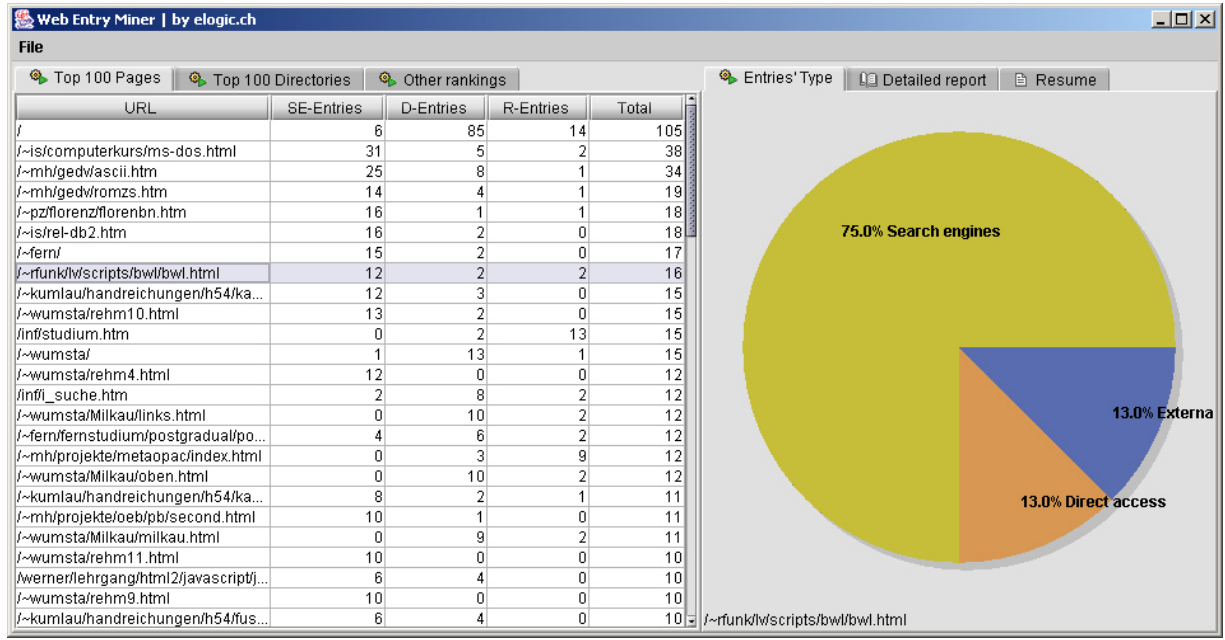

Figure 3: A visualization of a web entity (page) via the Java tool Web Entry Miner.

A next analysis step after the indicator computation (Tab. 1 or Fig. 3) could be the following 2-step scenario:

- Step 1: First the usage data will be analyzed with the proposed method. As a result we have a detailed macro view which distinguishes and displays the three usage aggregates, search engines, backlinks and direct access for each entity in a repository represented by their indicator values $I_{se}$, $I_{bl}$, $I_{da}$.
- Step 2: In the second step, the data analyst wishes to study the precise composition of the indicator values for a favorite entity, for example by its search engine indicator $I_{se}$. In doing so, this starts a "drill down" to the data (the downloads via search engines $D_{se}$ for the entity) which can display the exact configuration of the search engine accesses/downloads recorded in the log file. For example, 75% of search engine downloads refer back to the selected page in Figure 3 and these entries probably go back to several search engines and various search terms. The drill down analysis of the indicator can show all details, the search engines with the single search engine entry of the website referenced. In a further drill down step the search terms of an actual search engine, for example Google, will be shown. This is the point where the advanced macro analysis would end and a micro analysis could begin (see a micro-mining methodology in Nicholas & Huntington, 2003).

A drill-down scenario can also start with the backlink indicator $I_{bl}$. An analyst could be interested in the used backlinks/electronic citations of an analyzed document or entity. The distribution of backlinks can typically be viewed over time or by the intensity of the backlink use.

The following chapter will discuss the policy relevance and methodological difficulties of the proposed method. Furthermore we address the specific value and meaning of each indicator for evaluation of open accessible research documents.

## 4. Discussion

In accordance with the title of the paper, the proposed method to construct indicators for science and innovation systems out of usage data is highly experimental. But this experimental status is not only a specific property of our indicators, rather the status quo of most web-based indicators in these days is far from robust, including indicators which are build on linking data. Tendencies to develop these 'early' approaches to larger prototyping projects[5] can be observed. Usage data based methods (including our approach) have now little policy relevance for evaluators, because of some reasons. First there is the availability and privacy issues of the data. Normally raw or filtered usage data files are not available for everybody who is interested in the usage of a webserver or document repository. Especially the projects mentioned above (see footnote 5) show ideas/solutions how to overcome this problem. Secondly, the described method in this paper relies on the analysis of local log data, mostly only from one single webserver. Local log data is a limited data source in terms of measuring/evaluating science output and therefore it has only a little relevance for evaluators. A federated approach which has its own specific obstacles is a very promising direction with a higher impact for comprehensive science metrics. Evaluations of groups, institutes etc. would be better based on a federated approach. But we like to underline that our idea (distinguishing download pattern by a specific heuristic) has the potential to draw a more detailed picture of the usage statistics by establishing fine-grained indicators. Possibly this can be an add-on for enhanced evaluation of total counts of downloads. The third reason is the enormous amount of data and the problems in securing data quality/validity. Particularly the data quality and the validity of the constructed indicators should be mentioned here. Usage data with its absolutely urgency in filtering, extracting and validating of the raw data, could be relatively easily be manipulated, compared to citation or linking data. Log data analysis techniques are contentious, vague and full of traps (see e.g. Nicholas et al., 1999). Proxy server caching, user identification issues and other complicated problems like evaluating documents on distributed or mirrored systems by merging and sampling usage data are only some examples of problems you have to cope with, if you like to standardize such an evaluation method. All these reasons let us summarize that metrics which are based on usage data need an even better testing and circumspection before using it for evaluation or other interpretations. Furthermore this kind of metrics tend to be more successful in indicating/monitoring research output for user purposes. Users can better decide which publications are preferred by other users if downloads ratios are displayed. In accepting this shift from author/citation-centered to user/download-centered methodology, we have also a new and fast applicable basis for ranking and ordering search results or monitoring science trends with an increase in data vagueness.

The specific value and meaning of our approach lies in the detailed display of the total downloads of each single OA document or entities of items and the opportunity to dig deeper. The proposed indicators $I_{se}$, $I_{bl}$, $I_{da}$ open a microscopic way to analyze aggregates/instances of usage. They offer e.g. the possibility to discover hidden linking structures and usage pattern of groups. $I_{se}$, the search engine indicator, is a first indication for a collection manager if any open accessible document is indexed and retrieved from outside of the repository. A more detailed analysis of $I_{se}$ offer the possibility to learn more about the accessibility of items (see 2-step scenario in chapter 3). $I_{bl}$, the backlink indicator, is an immediate indication of impact of a document, because preference expressed by hyperlinks/'e-citations' and the usage of these links can be measured. Furthermore this indicator can show more than the absolute count of backlinks, but also the distribution of backlink usage over time. The $I_{bl}$ is conceptual related to traditional citation based indicators, because of the relatedness between citing and scholarly hyperlinking. It holds the greatest potential to build own/free forms of aggregation or validation with traditional metrics for evaluation purposes. $I_{da}$, the direct access indicator, is a specialty of our metric. It can indicate to which rate a document is downloaded directly.

---

[5] Projects to build indicators/metrics out of usage data: e.g. the bX project of the Los Alamos National Laboratory and Ex Libris or the Interoperable Repository Statistics (IRS) project by University of Southampton (http://irs.eprints.org/) presented recently (see footnote 3).

## 5. Conclusions

As we can show, logging data and derived web-based indicators of open accessible scholarly web spaces, is a very interesting source and also monitors real usage by real readers/users, in almost real time. Implementing widespread usage analysis and usage based indicators in OA repositories demonstrates benefits such as fast indication of usage impact and user preference. Today we are at a level where we should consider supplementing or enhancing traditional evaluation with new forms of web data. Web indicators based on usage data are a way to obtain unique information of the utility and scholarly acceptance of OA research output, but these new user oriented web indicators are still in an early experimental phase. The rationale of these approaches is strong but standardization of web metrics is at the beginning. In agreement with Thelwall's opinion that "providing information that may be used to take corrective action in research policy" (2004, p. 233) is an ambitious goal for our web-based indicators, research shows this to be a worthy goal, but there is still much to be done in evaluation of OA research. In addition we need further approaches to build robust measures and metrics especially for OA publications/documents. Following the thesis "science is turning to e-science" (Kretschmer & Aguillo, 2005; WISER website) we conclude e-science will need new stable e-indicators or a combination with traditional indicators to appropriately take e-only research output into consideration.

**Acknowledgement**

I would like to thank Jeof Spiro for helpful questions and proof reading of the manuscript.